\documentclass[12pt]{article}

\newcommand\eq[1] {(\ref{#1})}

\newcommand{\bfm}[1]{\mbox{\boldmath ${#1}$}}
\newcommand{\nonum}{\nonumber \\}
\newcommand{\beqa}{\begin{eqnarray}}
\newcommand{\eeqa}[1]{\label{#1}\end{eqnarray}}
\newcommand{\beq}{\begin{equation}}
\newcommand{\eeq}[1]{\label{#1}\end{equation}}
\newcommand{\Grad}{\nabla}
\newcommand{\Div}{\nabla \cdot}

\newcommand{\Real}{\mathop{\rm Re}\nolimits}
\newcommand{\Imag}{\mathop{\rm Im}\nolimits}

\newcommand{\Md}{\partial}

\newcommand{\Ga}{\alpha}
\newcommand{\Gb}{\beta}
\newcommand{\Gd}{\delta}

\newcommand{\Gve}{\varepsilon}

\newcommand{\Gg}{\gamma}

\newcommand{\Gn}{\eta}
\newcommand{\Gm}{\mu}
\newcommand{\Gv}{\nu}

\newcommand{\Gt}{\theta}

\newcommand{\BGb}{\bfm\beta}

\newcommand{\BGve}{\bfm\varepsilon}

\def\Bn{{\bf n}}

\def\BA{{\bf A}}

\def \ba {\begin{array}}
\def \ea {\end{array}}
\newtheorem {Thm} {Theorem} [section]

\newtheorem {Adef} [Thm] {Definition}

\newtheorem {Arem} [Thm] {Remark}

\newtheorem {Aexa} [Thm] {Example}

\newtheorem {Anot} [Thm] {Notation}

\def \refe #1.{(\ref{#1})}
\def \reff #1.{figure~\ref{#1}}
\def \refs #1.{section~\ref{#1}}
\def \refss #1.{subsection~\ref{#1}}
\def \refD #1.{Definition~\ref{#1}}
\def \refT #1.{Theorem~\ref{#1}}
\def \refL #1.{Lemma~\ref{#1}}
\def \refC #1.{Corollary~\ref{#1}}
\def \refP #1.{Proposition~\ref{#1}}
\def \refR #1.{Remark~\ref{#1}}
\def \refE #1.{Example~\ref{#1}}
\def \refN #1.{Notation~\ref{#1}}

\usepackage{graphics}
\usepackage{amssymb}
\begin{document}
\vspace{-1in}
\title{The searchlight effect in hyperbolic materials}
\author{Graeme W. Milton\\
\small{Department of Mathematics, University of Utah, Salt Lake City UT 84112, USA}\\
Ross C. McPhedran\\ 
\small{CUDOS, School of Physics, University of Sydney, NSW 2006, Australia}\\
Ari Sihvola\\
\small{Department of Radio Science and Engineering, Aalto University, 00076 Aalto, Finland}}
\date{}
\maketitle
\begin{abstract}
The quasistatic field around a circular hole in a two-dimensional hyperbolic medium is studied. As the loss parameter goes to zero, it is found that the electric field diverges along four lines each tangent to the hole. In this limit, the power dissipated by the field in the vicinity of these lines, per unit length of the line, goes to zero but extends further and further out so that the net power dissipated remains finite.  Additionally the interaction
between polarizable dipoles in a hyperbolic medium is studied. It is shown that a dipole with small polarizability can dramatically influence the
dipole moment of a distant polarizable dipole, if it is appropriately placed. We call this the searchlight effect, as the enhancement depends on the
orientation of the line joining the polarizable dipoles and can be varied by changing the frequency. For some particular polarizabilities the
enhancement can actually increase the further the polarizable dipoles are apart.  
\end{abstract}
\vskip2mm

\section{Introduction}
\setcounter{equation}{0}
Interest in hyperbolic materials, in which the dielectric tensor is real but with its eigenvalues taking different signs, has surged following the discovery of superlensing. The story of superlenses itself had its genesis in three pivotal papers.

The first by Veselago \cite{Veselago:1967:ESS} suggested that a slab of dielectric constant $\Gve=-1$ and magnetic permeability $\Gm=-1$ could act as a lens. This is indicated by ray tracing, taking into account the negative refractive index of the slab.

The second by
Nicorovici, McPhedran, and Milton \cite{Nicorovici:1994:ODP} (see also \cite{Milton:2005:PSQ}) showed that in two-dimensional quasistatics one could have apparent point singularities appear in the field surrounding a coated cylinder with coating having dielectric constant $\Gve=-1+i\Gd$, in the limit $\Gd\to 0$. Specifically, with $r_s$ and $r_c$ denoting the shell and core radii, and with a dipole source at $x=z_0$ outside the coated cylinder located in the annulus $a^2/r_s>z_0>r_s$ where $a=r_s^2/r_c$ it was proved in that paper that the complex potential $V(z)$ outside the coated cylinder, where $z=x+iy$ converged as $\Gd\to 0$ to the potential
\beq  \tilde{V}_e(z)=\frac{1}{z-z_0}-\left(\frac{1-\Gve_c}{1+\Gve_c}\right)\frac{a^2}{z_0^2(z-a^2/z_0)},
\eeq{0.1}
for all $z>a^2/z_0$, where $\Gve_c$ is the dielectric constant of the core. [The physical potential is  $\Real\{[V(z)+V(\overline{z})]e^{-i\omega t}/2\}$ where
$\omega$ is the frequency and $t$ is the time.] Thus an apparent point singularity appears at the point $z=a^2/z_0$ which {\it lies outside the coated cylinder}. In the limit $\Gd\to 0$, the shell acts to magnify the core by a factor of $r_s/r_c$ so it has the same response as a cylinder of radius $a$ 
and dielectric constant $\Gve_c$ (becoming invisible if $\Gve_c=1$) but now the "image dipole" lies in the matrix.  Within the radius 
$a^2/z_0>|z|>r_s$ it was numerically found that the potential develops enormous oscillations. This blowing up of the field within a localized region dependent on the position of the source, now called localized anomalous resonance,
may be physically regarded as a localized surface plasmon and is responsible for a type of invisibility cloaking \cite{Milton:2006:CEA, Nicorovici:2007:QCT},  
that has been the subject of considerable study \cite{Bruno:2007:SCS, Milton:2008:SFG, Nicorovici:2008:FWC, Bouchitte:2010:CSO, Nicorovici:2011:RLD, Ammari:2013:STN, Kohn:2012:VPC, Ammari:2013:STNII, Xiao:2012:TEC, Ammari:2013:ALR}.  

The third pivotal paper by Pendry \cite{Pendry:2000:NRM}, which served as a catalyst for the field, made the bold claim that the Veselago lens would be a superlens, capable of focussing much finer than the wavelength of the radiation. The appearance of apparent point singularities caused by localized anomalous resonance was later found to justify this claim for fixed amplitude point sources \cite{Ziolkowski:2001:WPM, Haldane:2002:ESM, Garcia:2002:LHM, Pokrovsky:2002:DLH, Cummer:2003:SCS, Pokrovsky:2003:DTF, Rao:2003:AEW, Shvets:2003:PAM, Merlin:2004:ASA, Guenneau:2005:PCR, Podolskiy:2005:NSS, Milton:2005:PSQ} though a single polarizable dipole is cloaked rather than perfectly imaged when it is sufficiently close to the superlens \cite{Milton:2006:CEA, Milton:2006:OPL, Xiao:2012:TEC}, and a small dielectric inclusion is at least partially cloaked \cite{Bruno:2007:SCS, Bouchitte:2010:CSO, Dong:2011:MSL}. Despite all the work on this topic (the paper of Pendry has over 6,700 citations) it remains an open question as to whether large dielectric inclusions (which interact with the surface plasmons)  are perfectly imaged when they are close to a superlens: work by Bruno and Lintner \cite{Bruno:2007:SCS}, would indicate they are not perfectly imaged (in the limit in which the loss in the lens goes zero)
while work of \cite{Dong:2011:MSL} suggests that they may be perfectly imaged, though it is not clear if sufficiently small loss has been taken in this latter investigation.

It was suggested by Pendry and Ramakrishna \cite{Pendry:2003:RPL} that a stack of layers of equal thicknesses alternating between $\Gve=+1$ and $\Gve=-1$,
would have an effective dielectric constant of infinity perpendicular to the layers and zero parallel to the layers, thus channelling the field like a set of infinitely conducting wires in an insulating matrix with $\Gve=0$. If the two constituent materials have different thicknesses and/or the dielectric constants with unequal magnitudes but opposite signs then the effective dielectric tensor can be a hyperbolic material (one needs to add a small loss to the material with negative dielectric constant to justify this, both physically and mathematically) and the dispersion relation in such materials allows for real wavevectors with arbitrarily large wavenumbers, thus allowing for propagation of waves with arbitrarily small wavelength \cite{Smith:2003:EWP}. The breakthrough came with the independent recognition of  Jacob, Alekseyev and 
Narimanov \cite{Jacob:2006:OHF} and Salandrino and Engheta \cite{Salandrino:2006:FFS}
that a multicoated cylinder or sphere with many thin coatings with dielectric constants of alternating signs would correspond to a hyperbolic material with radial symmetry
and be capable of magnifying an image from the subwavelength scale to a scale where conventional imaging would work: the hyperlens was born.
This was subsequently verified experimentally \cite{Liu:2007:FFO, Rho:2010:SHT}.

In this paper, our interest is in the two-dimensional quasistatic dielectric equation 
\beq \Div\BGve\Grad V=0,\quad   \BGve=\left[ \begin {array}{cc} \Gve_x  &  0 \\ \noalign{\medskip} 0 & \Gve_y \end {array} \right]. \eeq{0.2}
In the hyperbolic medium we consider $\Gve_x$ is real and $\Gve_y= \Gve_x/c^2$ with $c=-i\Gm+\Gn$ where $\Gn$ is a small positive parameter
($\Gm$ and $\Gn$ are real constants). In this medium, as $\Gn$ goes to zero, \eq{0.2} formally approaches the wave equation,
\beq \frac{\Md^2 V}{\Md y^2}=\Gm^2 \frac{\Md^2 V}{\Md x^2}, \eeq{0.3}
and thus in this limit one should expect wavelike solutions in the hyperbolic medium (think of $y$ as the time and $\Gm$ as the wave velocity). In this
paper we study how field singularities arise in the limit $\Gn\to 0$ when there is a circular hole in a hyperbolic material, and we study how a pair of
polarizable point dipoles interact in a hyperbolic medium when a uniform field is applied at infinity.

\section{The field around a circular hole in an anisotropic lossless medium}
\setcounter{equation}{0}
Here we review the known solution for the field surrounding a circular hole in a two-dimensional anisotropic medium, when the dielectric constants of the medium are real and positive. This problem has been treated before by Yang and Chou \cite{Yang:1977:ASP} in the context of the equivalent problem of antiplane elasticity, but their solution is more general than we need and they gave the potential only as an integral. In the next section we will use analytic continuation to obtain the solution when the dielectric constants of the medium are complex.  For this it is important to express the solution in cartesian coordinates rather than in stretched elliptical coordinates (which would be dependent on the dielectric constants of the medium and become unphysical when the dielectric constants of the medium become complex).

Consider the transformation
\beq z+\frac{r^2}{z}=2w, \eeq{1.1}
which maps the circle $|z|=r$ in the $z$ plane onto the slit $-r\leq w \leq r$ on the real $w$ axis, and maps a larger circle $|z|=r_0$, with $r_0>r$ to an
ellipse $E$, which in the $w=u+iv$ plane intersects the axes at
\beq u=\pm (r_0+r^2/r_0)/2,\quad v=\pm (r_0-r^2/r_0)/2. \eeq{1.2}
The transformation \eq{1.1} can be inverted to express $z$ and $1/z$ in terms of $w$:
\beq z=w+\sqrt{w^2-r^2},\quad \frac{1}{z}=\frac{w-\sqrt{w^2-r^2}}{r^2}.
\eeq{1.3}
It is to be emphasized that the square root needs to be taken so that the branch cut is a straight line between $w=r$ and $w=-r$: for computational purposes one can set
\beq \sqrt{w^2-r^2}=(w+r)\sqrt{(w-r)/(w+r)}, \eeq{1.3a}
where the square root on the right hand side is defined with the branch cut along the negative real axis. 
Let us suppose the material outside $E$ is isotropic with real positive dielectric constant $\Gve_x$, and now let us make some observations. First
consider the potential $\Gg w$ which at the surface $z=r_0 e^{i\Gt}$ equals
\beq \Gg(z+r^2/z)/2=\Gg(r_0 e^{i\Gt}+r^2e^{-i\Gt}/r_0)/2, \eeq{1.4}
and has complex conjugate
\beq \frac{\overline{\Gg}}{2}\left( \frac{r_0^2}{r_0 e^{i\Gt}}+\frac{r^2r_0e^{i\Gt}}{r_0^2}\right)= \overline{\Gg}t,
\eeq{1.5}
where $\overline{\Gg}$ is the complex conjugate of $\Gg$ and
\beqa t & = & \frac{1}{2}\left(\frac{r_0^2}{z}+\frac{r^2z}{r_0^2}\right) \nonum
& = & \frac{w}{2}\left(\frac{r^2}{r_0^2}+\frac{r_0^2}{r^2}\right)+\frac{\sqrt{w^2-r^2}}{2}\left(\frac{r^2}{r_0^2}-\frac{r_0^2}{r^2}\right).
\eeqa{1.6}
Thus the potential $\Real(\Gg w -\overline{\Gg}t)$ vanishes on the boundary $\partial E$, while the potential  $\Real(\Gg w+\overline{\Gg}t)$
has no flux of displacement field  across $\partial E$ (because the conjugate potential $\Imag(\Gg w+\overline{\Gg}t)$ vanishes on $\partial E$).

Let us take a parameter $c$ which to begin with we assume is real with $1>c>0$ and let us make the additional stretching transformation
\beq x=u,\quad y=v/c, \eeq{1.7}
which transforms the ellipse which intersects the $u$ and $v$ axes at the points \eq{1.2}
to an ellipse which intersects the $x$ and $y$ axes at the points
\beq x=\pm (r_0+r^2/r_0)/2,\quad y=\pm (r_0-r^2/r_0)/(2c). \eeq{1.8}
This final transformed ellipse will be the unit circle $x^2+y^2=1$ if we choose 
\beq r_0=1+c,\quad r=\sqrt{1-c^2}, \eeq{1.9}
so that
\beq (r_0+r^2/r_0)=2,\quad (r_0-r^2/r_0)=2c.
\eeq{1.10}
Inside this unit circle we put an isotropic medium with unit dielectric constant $\Gve_0=1$.

After this stretching transformation the dielectric tensor in the exterior medium becomes
\beq \BGve=\left[ \begin {array}{cc} \Gve_x  &  0 \\ \noalign{\medskip} 0 & \Gve_y \end {array} \right],
\eeq{1.10a}
where
\beq \Gve_y=\Gve_x/c^2.
\eeq{1.10b}
The potential $\Real(\Gg w -\overline{\Gg}t)$, expressed as a function of $x$ and $y$ will still vanish on the unit circle, while the potential  $\Real(\Gg w+\overline{\Gg}t)$
will still have no associated flux of displacement field across this boundary (assuming the dielectric tensor of the exterior medium is transformed to the anisotropic
value \eq{1.10a})
This trick of making an affine coordinate transformation to convert an isotropic matrix to an anistropic one has been used to find the solution
for an isotropic sphere in an anisotropic medium \cite{Sihvola:1997:DPI} and to find the effective dielectric tensor of assemblages of stretched
confocal coated ellipsoid assemblages (Section 8.4 of \cite{Milton:2002:TOC}). More generally it can be used to obtain explicit solutions
for the fields around three-dimensional ellipsoidal inclusions in a uniform applied field with an anisotropic core and anisotropic matrix, each with arbitrary orientation. 

Now consider a potential $V(x,y)$ given by
\beqa V(x,y) & = & \Real(\Gb w +\overline{\Gg} t)\quad{\rm for}\quad x^2+y^2\geq 1,
\nonum
                   & = & \Gd_x x +\Gd_y y\quad{\rm for}\quad x^2+y^2<1,
\eeqa{1.11}
where $w=u+iv=x+icy$ and from \eq{1.6} and \eq{1.9} $t$ is given by 
\beq t =\frac{(1+c^2)(x+icy)}{1-c^2}+\frac{2c\sqrt{(x+icy)^2+c^2-1}}{c^2-1}.
\eeq{1.11a}
At the boundary of the unit circle $\Real(\overline{\Gg} t)=\Real(\Gg w)$ so continuity of the potential $V(x,y)$ requires
\beq \Real[(\Gb+\Gg)w]= \Gd_x x +\Gd_y y.
\eeq{1.12}
Also continuity of the normal component of the displacement field requires
\beq n_x\Gve_x\frac{\Md}{\Md x} \Real(\Gb w +\overline{\Gg} t)+ n_y\Gve_y\frac{\Md}{\Md y} \Real(\Gb w +\overline{\Gg} t)=n_x\Gd_x+n_y\Gd_y,
\eeq{1.13}
where $\Bn=(n_x,n_y)=(x/\sqrt{x^2+y^2}, y/\sqrt{x^2+y^2})$ is the unit outward normal to the boundary of the unit disk. 
Since the potential  $\Real(\Gg w+\overline{\Gg}t)$ has no associated flux of displacement field across this boundary we have 
\beq n_x\Gve_x\frac{\Md}{\Md x} \Real(\overline{\Gg} t)+ n_y\Gve_y\frac{\Md}{\Md y} \Real(\overline{\Gg} t)=
n_x\Gve_x\frac{\Md}{\Md x} \Real(-\Gg w)+ n_y\Gve_y\frac{\Md}{\Md y} \Real(-\Gg w),
\eeq{1.14}
so the flux continuity condition \eq{1.13} reduces to
\beq n_x\Gve_x\frac{\Md}{\Md x} \Real[(\Gb-\Gg)w]+ n_y\Gve_y\frac{\Md}{\Md y} \Real[(\Gb-\Gg)w]=n_x\Gd_x+n_y\Gd_y.
\eeq{1.15}
When $\Gb$ and $\Gg$ are real, $\Gb=\Gb'$ and $\Gg=\Gg'$, corresponding to an applied field acting in the $x$-direction, then $\Gd_y=0$ and \eq{1.12} and \eq{1.15} are satisfied when
\beq \Gb'+\Gg'= \Gd_x, \quad \Gve_x(\Gb'-\Gg')=\Gd_x.
\eeq{1.16}
These have the solution
\beq \Gd_x=\frac{2\Gg'\Gve_x}{\Gve_x-1}, \quad \Gb'=\frac{\Gg'(\Gve_x+1)}{\Gve_x-1},
\eeq{1.17}
and from \eq{1.11} and \eq{1.9} the potential outside the inclusion,
\beq V(x,y)  =  \Gb' x +\Gg'\Real(t), 
\eeq{1.17a}
equates to
\beqa
V(x,y) & = & \frac{\Gg'(\Gve_x+1)x}{\Gve_x-1}+\frac{\Gg' (1+c^2)x}{1-c^2}\nonum
& & +\frac{\Gg'c\left[\sqrt{(x+icy)^2+c^2-1}+\sqrt{(x-icy)^2+c^2-1}\right]}{c^2-1},
\eeqa{1.18}
while the field inside is
\beq V(x,y)=\Gd_x x=\frac{2\Gg'\Gve_x x}{\Gve_x-1}.
\eeq{1.19}

On the other hand when $\Gb$ and $\Gg$ are imaginary, $\Gb=i\Gb''$ and $\Gg=i\Gg''$, corresponding to an applied field acting in the $y$-direction, then $\Gd_x=0$ and \eq{1.12} and \eq{1.15} are satisfied when
\beq -c(\Gb''+\Gg'')= \Gd_y, \quad -\Gve_y c(\Gb''-\Gg'')=\Gd_y.
\eeq{1.20}
These have the solution
\beq \Gd_y=\frac{2c\Gg''\Gve_y}{1-\Gve_y}, \quad \Gb''=\frac{\Gg''(\Gve_y+1)}{\Gve_y-1},
\eeq{1.21}
and from \eq{1.11} and \eq{1.9} the potential outside the inclusion,
\beq V(x,y) =  -\Gb''cy+\Gg''\Imag(t), \eeq{1.21a}
equates to
\beqa  V(x,y)  & = & \frac{\Gg''(\Gve_y+1)cy}{1-\Gve_y}+\frac{\Gg'' (1+c^2)cy}{1-c^2}\nonum
& & +\frac{\Gg''c\left[\sqrt{(x+icy)^2+c^2-1}-\sqrt{(x-icy)^2+c^2-1}\right]}{i(c^2-1)},
\eeqa{1.22}
while the field inside is
\beq V(x,y)=\Gd_y y=\frac{2c\Gg''\Gve_y y}{1-\Gve_y}.
\eeq{1.23}

\section{The field singularities around a circular hole in a hyperbolic medium}
\setcounter{equation}{0}
The potential $V(x,y)$ given by \eq{1.18} and \eq{1.19}, or by \eq{1.22} and \eq{1.23}, solves the dielectric equations when $c$ and $\Gve_x$ take any real positive value. By analytic continuation these formulae also solve the dielectric equations when $c$ and $\Gve_x$  are complex, and in this case $V(x,y)$ is given by substituting the complex values of $c$, $\Gve_x$ and  $\Gve_y=\Gve_x/c^2$ in these formula. The branch cuts in the square roots need to be taken so (for fixed non-zero $\Gn$) there are no singularities in the field outside the cylinder. Guided by \eq{1.3a} we choose
\beq \sqrt{(x+icy)^2+c^2-1} = (x+icy+\sqrt{1-c^2})\sqrt{\frac{x+icy-\sqrt{1-c^2}}{x+icy+\sqrt{1-c^2}}}, \eeq{2.0a}
\beq  \sqrt{(x-icy)^2+c^2-1} = (x-icy+\sqrt{1-c^2})\sqrt{\frac{x-icy-\sqrt{1-c^2}}{x-icy+\sqrt{1-c^2}}}, \eeq{2.0b}
where the square roots on the right hand side of these expressions have their branch cuts along the negative real axis. The expression beneath the square root
in \eq{2.0a} will be real when
\beq \frac{x+icy-\sqrt{1-c^2}}{x+icy+\sqrt{1-c^2}}=\frac{x-i\overline{c}y-\overline{\sqrt{1-c^2}}}{x-i\overline{c}y+\overline{\sqrt{1-c^2}}},
\eeq{2.0c}
(where the bar denotes complex conjugation) which is satisfied when $(x,y)$ lies on the line
\beq y\Real[\overline{c}\sqrt{1-c^2}]=x\Imag[\sqrt{1-c^2}].
\eeq{2.0d}
Along this line the ratio in \eq{2.0c} will be negative along the interval between the points where $x+icy =\pm \sqrt{1-c^2}$, i.e. between the points
\beq (x,y)=\pm\left(\frac{\Real[\overline{c}\sqrt{1-c^2}]}{\Real[c]},\frac{\Imag[\sqrt{1-c^2}]}{\Real[c]}\right), \eeq{2.0e}
which will lie in the unit disk $x^2+y^2<1$ if and only if
\beq \left(\Real[c]\right)^2-\left(\Real[\overline{c}\sqrt{1-c^2}]\right)^2-\left(\Imag[\sqrt{1-c^2}]\right)^2>0.
\eeq{2.0f}
Numerically we have checked that the left hand side is always non-negative and zero only when $c$ is purely imaginary. This shows that there are no branch cuts in the potential outside the circular hole when $\Gn>0$.

The parameters  $\Gg'$, $\Gg''$,  $\Gb'$, $\Gb''$, $\Gd_x$, $\Gd_y$, $r$ and $r_0$ are also generally complex.
Let us take $\Gve_x$ to be real and positive and $c=-i\Gm+\Gn$ where $\Gn$ is a small positive parameter. Then
\beq \Gve_y=\Gve_x/(-i\Gm+\Gn)^2\approx -\Gve_x/\Gm^2+2i\Gve_x\Gn/\Gm^3
\eeq{2.1}
is close to being real and negative. Since
\beqa \sqrt{(x+icy)^2+c^2-1} & = & \sqrt{(x+\Gm y+i\Gn y+\sqrt{1-c^2})(x+\Gm y+i\Gn y-\sqrt{1-c^2})}, \nonum
\sqrt{(x-icy)^2+c^2-1} & = & \sqrt{(x-\Gm y-i\Gn y+\sqrt{1-c^2})(x-\Gm y-i\Gn y-\sqrt{1-c^2})},
\eeqa{2.2}
we see that the potential $V(x,y)$ given by \eq{1.18} or \eq{1.22} develops singularities as $\Gn\to 0$ along the four characteristic lines
\beqa x+\Gm y+\sqrt{1+\Gm^2} & = & 0,\quad x+\Gm y-\sqrt{1+\Gm^2}=0,\nonum x-\Gm y+\sqrt{1+\Gm^2}& = & 0, \quad x-\Gm y-\sqrt{1+\Gm^2}=0,
\eeqa{2.3}
which are tangent to the unit disk touching it at the four points 
\beq (x,y)=(\pm 1/\sqrt{1+\Gm^2},\pm \Gm/\sqrt{1+\Gm^2}).
\eeq{2.4}
Figure 1 shows a plot of the absolute value of the potential  $V(x,y)$ showing how kinks develop along the characteristic lines

\begin{figure}[htbp]
\begin{center}
\scalebox{0.5}{\includegraphics{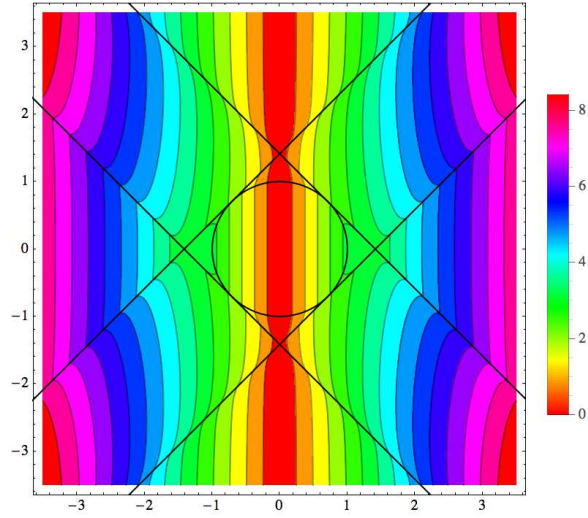}} 
\end{center}
\caption{Plot of the absolute value of the potential $V(x,y)$ around the disk given by \protect{\eq{1.18}} with an applied field directed along the $x$-axis and with parameters $\Gg'=1$, $\Gve_x=3$ and $c=0.01-i$. The potential has kinks at the four characteristic lines given by \protect{\eq{2.3}} which are also drawn. Near these lines
the electric field is huge.}
\label{fig:1}
\end{figure}

To better understand the behavior of the fields near these singularities let us consider the potential $V(x,y)$ given by \eq{1.18} in a region near the characteristic line
 $x+\Gm y+\sqrt{1+\Gm^2}=0$ but away from the disk and away from the three other characteristic lines. In this region $V(x,y)$ takes the form
\beq V(x,y)=H(x,y)+G(x,y)\sqrt{x+\Gm y+i\Gn y+r}, \eeq{2.5}
where 
\beqa r & = & \sqrt{1-c^2}\approx \sqrt{1+\Gm^2}+\frac{i\Gm\Gn}{\sqrt{1+\Gm^2}}, \nonum
H(x,y) & = & \frac{\Gg'(\Gve_x+1)x}{\Gve_x-1}+\frac{\Gg' (1+c^2)x}{1-c^2}+\frac{\Gg'c\sqrt{(x-icy)^2-r^2}}{c^2-1}, \nonum
G(x,y) & = & \frac{\Gg'c\sqrt{x+icy-r}}{c^2-1}.
\eeqa{2.6}
Near the characteristic line the electric field blows up as $\Gn\to 0$ and
\beq \frac{\Md V}{\Md y} \approx \frac{\Gm G_0}{\sqrt{x+\Gm y+i\Gn y+r}}\approx\frac{\Gm G_0}{\sqrt{h+i\Gn s}},
\eeq{2.7}
where
\beq G_0 =  \frac{\Gg' i\Gm \sqrt{-2\sqrt{1+\Gm^2}}}{1+\Gm^2}=-\Gg'\Gm\sqrt{2}(1+\Gm^2)^{-3/4}
\eeq{2.8}
is the limit as $\Gn\to 0$ of $G(x,y)$ on the line $x+\Gm y+\sqrt{1+\Gm^2}=0$, and
where
\beq h=x+\Gm y+\sqrt{1+\Gm^2}, \quad s=y+\frac{\Gm}{\sqrt{1+\Gm^2}}.
\eeq{2.9}
(Thus $h$ measures the distance from the characteristic line  $x+\Gm y+\sqrt{1+\Gm^2}=0$.) 
So the local time averaged power dissipated in this region per unit area is proportional to 
\beqa \Imag(\Gve_y)\left|\frac{\Md V}{\Md y}\right|^2 & \approx &  \frac{|G_0|^2\Gm^2 \Imag(\Gve_y)}{|h+i\Gn s|} \nonum
&\approx& \frac{2\Gn\Gve_x|G_0|^2}{\Gm\sqrt{h^2+\Gn^2s^2}},
\eeqa{2.10}
in which \eq{2.1} has been used to estimate $\Imag(\Gve_y)$. Let us change variables from $(x,y)$ to $(h,s)$, so that $dx\,dy=dh\,ds$. Observe that the right side of
\eq{2.9} is an even function of $h$ and that the integral
\beq \int_0^{h_0}\frac{dh}{\sqrt{h^2+\Gn^2s^2}}=\ln(h_0+\sqrt{h^2+\Gn^2s^2})-\ln(\Gn|s|),
\eeq{2.11}
when $\Gn$ is very small, and $h_0$ is not too large (so the approximation of being near the characteristic line is still valid) has a dominant contribution of $-\ln(\Gn|s|)$. So when $\Gn$ is very small we have
\beq \int\Imag(\Gve_y)\left|\frac{\Md V}{\Md y}\right|^2\,dh\approx -4\Gve_x|G_0|^2\Gn\ln(\Gn|s|).
\eeq{2.12}
Hence along the characteristic line the power absorption (integrated across the line), per unit length of the characteristic line, goes as $\Gn\ln(\Gn|s|)$ which goes to zero as $\Gn\to 0$. However it
extends a long way out along these characteristic lines so the total contribution does not tend to zero. To see this first note that the approximation \eq{2.7} will clearly break down at large values of $s$, specifically when $\Gn s$ is of the order of one, since then the right hand side of \eq{2.7} becomes comparable to $\Md H/\Md y$. 
Thus a ball park estimate of the total absorption coming from this characteristic line is
\beq 2\int_0^{1/\Gn} -4\Gve_x|G_0|^2\Gn\ln(\Gn|s|)\,ds=8\Gve_x|G_0|^2. \eeq{2.13}
The interesting point is that this total absorption remains finite and non-zero as $\Gn\to 0$. 
This explains the discovery of Sihvola \cite{Sihvola:2005:MDF} that a hole in a hyperbolic medium 
may have loss even though the medium is essentially lossless. 
 
\section{The dipole approximation for the far field  around a circular hole in a hyperbolic medium}
\setcounter{equation}{0}
Consider the potential $V(x,y)$ given by equation \eq{1.18} corresponding to an applied field in the $x$-direction. When $x+icy$ and $x-icy$ are both large we can use the approximations
\beqa \sqrt{(x+icy)^2+c^2-1}& \approx &  x+icy+\frac{c^2-1}{2(x+icy)}, \nonum
 \sqrt{(x-icy)^2+c^2-1}& \approx &  x-icy+\frac{c^2-1}{2(x-icy)},
\eeqa{3.1}
to obtain 
\beqa V(x,y) 
& \approx & \frac{\Gg'(\Gve_x+1)x}{\Gve_x-1}+\frac{\Gg' (1+c^2)x}{1-c^2}
 -\frac{2\Gg'c x}{1-c^2}+\frac{\Gg'c}{2(x+icy)}+\frac{\Gg'c}{2(x-icy)} \nonum
&\approx & \Gg'x\left(\frac{\Gve_x+1}{\Gve_x-1}+\frac{1-c}{1+c}\right)
+\frac{\Gg'c x}{x^2+c^2y^2} \nonum
&\approx &  \Gg'\left(\frac{\Gve_x+1}{\Gve_x-1}+\frac{1-c}{1+c}\right)\left(x
-\frac{x\Ga_x}{x^2+c^2y^2}\right), 
\eeqa{3.2}
where $\Ga_x$ is the polarizability
\beq \Ga_x= \frac{-c}{\left(\frac{\Gve_x+1}{\Gve_x-1}+\frac{1-c}{1+c}\right)}=\frac{c(1+c)(1-\Gve_x)}{2(\Gve_x+c)},
\eeq{3.3}
which has been normalized to make the last bracketed expression in \eq{3.2} as simple as possible.

Similarly, when the applied field is in the $y$ direction and $x+icy$ and $x-icy$ are both large the potential given by \eq{1.22} has the far field behavior
\beqa V(x,y) 
& \approx & \frac{\Gg''(\Gve_y+1)cy}{1-\Gve_y}+\frac{\Gg'' (1+c^2)cy}{1-c^2}
 -\frac{2\Gg''c^2 y}{1-c^2}+\frac{\Gg''c}{2i(x+icy)}-\frac{\Gg''c}{2i(x-icy)} \nonum
&\approx & \Gg''cy\left(\frac{\Gve_y+1}{1-\Gve_y}+\frac{1-c}{1+c}\right)
-\frac{\Gg''c^2 y}{x^2+c^2y^2} \nonum
&\approx &  \Gg''cy\left(\frac{\Gve_y+1}{1-\Gve_y}+\frac{1-c}{1+c}\right)\left(y
-\frac{y\Ga_y}{x^2+c^2y^2}\right),
\eeqa{3.4}
where $\Ga_y$ is the normalized polarizability
\beq \Ga_y=\frac{c}{\left(\frac{\Gve_y+1}{1-\Gve_y}+\frac{1-c}{1+c}\right)}=\frac{c(1+c)(1-\Gve_y)}{2(1+c\Gve_y)}.
\eeq{3.5}
We call \eq{3.2} and \eq{3.3} the dipole approximation for the far field. 
Note that in a hyperbolic medium it does not suffice for $x^2+y^2$ to be large to ensure that both  $x+icy$ and $x-icy$ are large. One must also be sufficiently
distant from the lines $x=\pm \Gm y$ since along these lines either $x+icy$ or $x-icy$ is close to zero when $\Gn$ is small. Thus for the dipole approximation
for the far field to be valid one must be sufficiently far from the four characteristic lines \eq{2.3}: this makes sense as the electric field diverges to infinity along these lines, whereas the dipole field only diverges on the two lines $x=\pm \Gm y$ as $\Gn\to 0$.

The expressions \eq{3.3} and \eq{3.5} for the polarizabilities could have been obtained more easily from the far field expressions for
the potential outside an elliptical hole in an isotropic medium. When $c$ is real and positive and before the stretching, the elliptical hole has axis
lengths of $2$ and $2c$ and in the $(u,v)$ plane. With an applied field in the $u$-direction the far field in the isotropic medium with
dielectric constant $\Gve_x$ has potential
\beq u-\frac{\pi c u(1-\Gve_x)}{2\pi(u^2+v^2)\left[\Gve_x+(1-\Gve_x)\frac{c}{1+c}\right]}, \eeq{3.6}
where $\pi c$ is the area of the ellipse and $c/(1+c)$ is the depolarization factor of the ellipse in the $u$-direction. By making the transformation
$x=u$, $y=v/c$ this potential gets mapped to
\beq x-\frac{x\Ga_x}{x^2+c^2y^2}, \eeq{3.7}
where the polarizability $\Ga_x$ is given by \eq{3.3}. It is similarly easy to derive \eq{3.5} using the depolarization factor $1/(1+c)$ of the ellipse in the $v$ direction. When $c=1$ and $\Gve_y=\Gve_x$ \eq{3.3} and \eq{3.5} reduce to 
$\Ga_x=\Ga_y=(1-\Gve_x)/(1+\Gve_x)$ which (within a proportionality factor) is the polarizability of a hole in an isotropic medium having
dielectric constant $\Gve_x$.

\section{The response of two interacting polarizable dipoles in a hyperbolic medium}
\setcounter{equation}{0}
Here we study the mathematics of the interaction of two ideal polarizable dipoles in a hyperbolic medium. We leave open the question as to whether these ideal polarizable dipoles have any physical significance. 
Nevertheless the searchlight effect discussed here should motivate future studies to see, say, whether a distant very small circular disk, if appropriately positioned, can substantially influence the response of a large circular disk.

By definition a polarizable dipole with rectangular symmetry located at the origin responds to a local field acting on it in the $x$ direction so that the potential $V(x,y)$ close to the origin has the expansion 
\beq V(x,y)\approx x+a_1-\frac{\Ga_x/2}{x+icy}-\frac{\Ga_x/2}{x-icy}=x+a_1-\frac{x\Ga_x}{x^2+c^2y^2}, \eeq{4.1}
and responds to an local field acting on it in the $y$ direction so that the potential $V(x,y)$ close to the origin has the expansion
\beq V(x,y) \approx y+a_2+\frac{\Ga_y/(2ic)}{x+icy}-\frac{\Ga_y/(2ic)}{x-icy}=y+a_2-\frac{y\Ga_y}{x^2+c^2y^2}. \eeq{4.2}
Here we call $\Ga_x$ and $\Ga_y$ the polarizabilities of the polarizable dipole, $a_1$ and $a_2$ are constants, and $c=i\Gm-\Gn$ with $\Gn$ being small. By taking linear combinations, the response to an arbitrarily oriented local field is such that the potential close to the origin has the expansion
\beq V(x,y)\approx \Gg_x x+\Gg_y y+a-\frac{x\Gg_x\Ga_x+y\Gg_y\Ga_y}{x^2+c^2y^2}. \eeq{4.3}

With $c=-i\Gm+\Gn$ and $\Gn>0$ small, the potential on the right hand side of \eq{4.1} now has a local time averaged power dissipation near the
characteristic line $x+\Gm y=0$ of
\beqa  \Imag(\Gve_y)\left|\frac{\Md V}{\Md y}\right|^2 & \approx &  \frac{\Gm^2|\Ga_x|^2\Imag(\Gve_y)}{4|g+i\Gn y|^4} \nonum
&\approx& \frac{\Gn\Gve_x|\Ga_x|^2}{2\Gm(g^2+\Gn^2y^2)^2},
\eeqa{4.3a}
where $g=x+\Gm y$. So when $\Gn$ is very small the dissipation integrated with respect to $g$, in the range $-g_0\geq g\geq g_0$, 
is approximately
\beq \int_{-g_0}^{g_0}  \Imag(\Gve_y)\left|\frac{\Md V}{\Md y}\right|^2\,dg\approx
\frac{\Gve_x|\Ga_x|^2}{2\Gm y^3\Gn^2}\int_{-\infty}^{\infty}\frac{d\Gv}{(\Gv^2+1)^2},
\eeq{4.3b}
where $\Gv=g/(\Gn y)$. Thus this power dissipation integrated across the characteristic line, per unit length of the characteristic line, blows
up as $\Gn\to 0$.

Now consider a uniform applied field in the $x$-direction acting on two polarizable dipoles, each with rectangular symmetry, one located at the origin and the
other at the point $(x_0,y_0)$.  
The total field $V(x,y)$ is the sum of the uniform field plus the two dipolar fields:
\beq V(x,y)=x+ \frac{x\beta_{1x}+y\beta_{1y}}{x^2+c^2y^2}+\frac{(x-x_0)\beta_{2x}+(y-y_0)\beta_{2y}}{(x-x_0)^2+c^2(y-y_0)^2}.
\eeq{4.10}
Expanding this around the origin $x=y=0$ gives
\beqa V(x,y)&\approx & \frac{x\beta_{1x}+y\beta_{1y}}{x^2+c^2y^2}-\frac{(x_0\beta_{2x}+y_0\beta_{2y})}{x_0^2+c^2y_0^2} \nonum
& & +x+\frac{x\beta_{2x}+y\beta_{2y}}{x_0^2+c^2y_0^2}
-\frac{(2xx_0+2yy_0)(x_0\beta_{2x}+y_0\beta_{2y})}{(x_0^2+c^2y_0^2)^2},
\eeqa{4.11}
which allows us to identify the local field acting on the dipole at the origin. 
If $(\alpha_{1x},\alpha_{1y})$ are the polarizability coefficients of the dipole at the origin then from \eq{4.3} we have
\beqa \beta_{1x}& = & \left[ -1-\frac{\beta_{2x}}{x_0^2+c^2y_0^2}+\frac{2x_0(x_0\beta_{2x}+y_0\beta_{2y})}{(x_0^2+c^2y_0^2)^2}\right]\alpha_{1x}, \nonum
 \beta_{1y}& = & \left[ -\frac{\beta_{2y}}{x_0^2+c^2y_0^2}+\frac{2c^2y_0(x_0\beta_{2x}+y_0\beta_{2y})}{(x_0^2+c^2y_0^2)^2}\right]\alpha_{1y}.
\eeqa{4.12}
In a similar fashion, by rewriting \eq{4.10} as
\beqa V(x,y)=x_0+(x-x_0)+ \frac{[x_0+(x-x_0)]\beta_{1x}+[y_0+(y-y_0)]\beta_{1y}}{[x_0+(x-x_0)]^2+c^2[y_0+(y-y_0)]^2}+\frac{(x-x_0)\beta_{2x}+(y-y_0)\beta_{2y}}{(x-x_0)^2+c^2(y-y_0)^2}, \nonum
\eeqa{4.13}
and expanding this around the point $(x_0,y_0)$ we obtain
\beqa V(x,y) & \approx & \frac{(x-x_0)\beta_{2x}+(y-y_0)\beta_{2y}}{(x-x_0)^2+c^2(y-y_0)^2}
+x_0+\frac{(x_0\beta_{1x}+y_0\beta_{1y})}{x_0^2+c^2y_0^2} \nonum
& &+(x-x_0)+\frac{(x-x_0)\beta_{1x}+(y-y_0)\beta_{1y}}{x_0^2+c^2y_0^2} \nonum
& &-\frac{[2(x-x_0)x_0+2(y-y_0)y_0](x_0\beta_{1x}+y_0\beta_{1y})}{(x_0^2+c^2y_0^2)^2}.
\eeqa{4.14}
So if  $(\alpha_{2x},\alpha_{2y})$ are the polarizability coefficients of the dipole at the point $(x_0,y_0)$ then we have
\beqa \beta_{2x}& = & \left[ -1-\frac{\beta_{1x}}{x_0^2+c^2y_0^2}+\frac{2x_0(x_0\beta_{1x}+y_0\beta_{1y})}{(x_0^2+c^2y_0^2)^2}\right]\alpha_{2x}, \nonum
 \beta_{2y}& = & \left[ -\frac{\beta_{1y}}{x_0^2+c^2y_0^2}+\frac{2c^2y_0(x_0\beta_{1x}+y_0\beta_{1y})}{(x_0^2+c^2y_0^2)^2}\right]\alpha_{2y}.
\eeqa{4.15}
Introducing
\beq e=x_0^2+c^2y_0^2 \approx x_0^2-\Gm^2y_0^2-2i\Gm\Gn y_0^2, \eeq{4.16}
in terms of which $c^2=(e-x_0^2)/y_0^2$, the equations \eq{4.12} and \eq{4.15} take the equivalent form
\beqa \frac{\beta_{1x}}{\alpha_{1x}} & = & -1-\frac{\beta_{2x}}{e}+\frac{2x_0(x_0\beta_{2x}+y_0\beta_{2y})}{e^2},\nonum
 \frac{\beta_{1y}}{\alpha_{1y}}& = & -\frac{\beta_{2y}}{e}+\frac{2(e-x_0^2)(x_0\beta_{2x}+y_0\beta_{2y})}{y_0e^2}, \nonum
 \frac{\beta_{2x}}{\alpha_{2x}}& = & -1-\frac{\beta_{1x}}{e}+\frac{2x_0(x_0\beta_{1x}+y_0\beta_{1y})}{e^2},\nonum
 \frac{\beta_{2y}}{\alpha_{2y}}& = & -\frac{\beta_{1y}}{e}+\frac{2(e-x_0^2)(x_0\beta_{1x}+y_0\beta_{1y})}{y_0e^2}.
\eeqa{4.17}
These four equations have the solution
\beq \BGb\equiv\left[ \begin {array}{c} \beta_{1x} \\ \noalign{\medskip} \beta_{1y}
\\ \noalign{\medskip}\beta_{2x}\\ \noalign{\medskip}\beta_{2y}\end {array} \right]= \BA^{-1}
\left[ \begin {array}{c} -1\\ \noalign{\medskip}0
\\ \noalign{\medskip}-1\\ \noalign{\medskip}0\end {array} \right],
\eeq{4.18}
where $\BA$ is the matrix
\beq \BA=\left[ \begin {array}{cccc} \frac{1}{\alpha_{{1x}}}&0&\frac{1}{e}-{
\frac {2\,{x_{{0}}}^{2}}{{e}^{2}}}&-{\frac {2\,x_{{0}}y_{{0}}}{{e}^{2}}}
\\ \noalign{\medskip}0&\frac{1}{\alpha_{{1y}}}&-{\frac {2\,x_{{0}}
 \left( e-{x_{{0}}}^{2} \right) }{{e}^{2}y_{{0}}}}&
\quad\,-\frac{1}{e}+{
\frac {2\,{x_{{0}}}^{2}}{{e}^{2}}}
\\ \noalign{\medskip}\frac{1}{e}-{
\frac {2\,{x_{{0}}}^{2}}{{e}^{2}}}&-{\frac {2\,x_{{0}}y_{{0}}}{{e}^{2}}}&
\frac{1}{\alpha_{{2x}}}&0\\ \noalign{\medskip}-{\frac {2\,x_{{0}} \left( e
-{x_{{0}}}^{2} \right) }{{e}^{2}y_{{0}}}}&\quad\,-\frac{1}{e}+{
\frac {2\,{x_{{0}}}^{2}}{{e}^{2}}}
&0&\frac{1}{\alpha_{{2y}}}\end {array} \right],
\eeq{4.18a}
which has determinant
\beq {\rm det}(\BA)=\frac{\left[4\,{x_{{0}}}^{2}(e-{x_{{0}}}^{2})(\alpha_{{1x}}-\alpha_{{1y}})(\alpha_{{2x}}-\alpha_{{2y}})+ 
\left( {e}^{2}-\alpha_{{1x}}\alpha_{{2x}} \right)  \left( {e}^{2}-\alpha_{{1y}}\alpha_{{2y}} \right)\right]}{e^4\alpha_{{1x}}\alpha_{{1y}}\alpha_{{2x}}\alpha_{{2y}}},
\eeq{4.18b}
that vanishes when $x_0^2$ solves the quadratic
\beq 4\,{x_{{0}}}^{2}(e-{x_{{0}}}^{2})(\alpha_{{1x}}-\alpha_{{1y}})(\alpha_{{2x}}-\alpha_{{2y}})+ 
\left( {e}^{2}-\alpha_{{1x}}\alpha_{{2x}} \right)  \left( {e}^{2}-\alpha_{{1y}}\alpha_{{2y}} \right)=0.
\eeq{4.18c}

We are interested in what happens to this solution when $x_0$ and $y_0$ are such that $e$ is very small. Using Maple to compute the matrix inverse and taking the
limit $e\to 0$ we find that
\beq \BGb\to
 \left[ \begin {array}{c} {\frac {2\,{x_{{0}}}^{2}\alpha_{{1x}}\alpha_{
{1y}} \left[ 2\,{x_{{0}}}^{2}(\alpha_{{2y}}-\alpha_{{2x}})+
\alpha_{{2y}}\alpha_{{2x}} \right] }{\alpha_{{1y}}\alpha_{{2y}}\alpha_{{1x}}
\alpha_{{2x}}-4\,{x_{{0}}}^{4}(\alpha_{{1x}}-\alpha_{{1y}})(\alpha_{{2x}}-\alpha_{{2y}})}}\\ \noalign{\medskip}-{\frac 
{2\,{x_{{0}}}^{3}\alpha_{{1x}}\alpha_{{1y}} \left[ 2\,{x_{{0}}}^{2}
(\alpha_{{2y}}-\alpha_{{2x}})+\alpha_{{2y}}\alpha_{{2x}} \right] }{y_{{0
}} \left[ \alpha_{{1y}}\alpha_{{2y}}\alpha_{{1x}}\alpha_{{2x}}-4\,{x_{{0}}}^{4}(\alpha_{{1x}}-\alpha_{{1y}})(\alpha_{{2x}}-\alpha_{{2y}})
\right] }}\\ \noalign{\medskip}{\frac {2\,{x_{{0}}}^{2}\alpha_{{2
x}}\alpha_{{2y}} \left[ 2\,{x_{{0}}}^{2}(\alpha_
{{1y}}-\alpha_{{1x}})+\alpha_{{1x}}\alpha_{{1y}} \right] }{\alpha_{{1y}}\alpha_{{2y}}
\alpha_{{1x}}\alpha_{{2x}}-4\,{x_{{0}}}^{4}(\alpha_{{1x}}-\alpha_{{1y}})(\alpha_{{2x}}-\alpha_{{2y}})}}\\ \noalign{\medskip}-
{\frac {2\,{x_{{0}}}^{3}\alpha_{{2y}}\alpha_{{2x}} \left[ 2\,{x_{{0}}}^{2}(\alpha_{{1y}}-\alpha_{{1x}})+\alpha_{{1x}}\alpha_{{1y}}
 \right] }{y_{{0}} \left[ \alpha_{{1y}}\alpha_{{2y}}\alpha_{{1x}}\alpha_{{2x
}}-4\,{x_{{0}}}^{4}(\alpha_{{1x}}-\alpha_{{1y}})(\alpha_{{2x}}-\alpha_{{2y}})
\right] }}\end {array} \right].
\eeq{4.19}
Thus when  $\alpha_{1x}\ne \alpha_{1y}$ and  $\alpha_{2x}\ne \alpha_{2y}$  we see that in the limit $e\to 0$ all components of $\Gb$ blow up to infinity when $x_0$ is such that
\beq x_0^4=\frac{\alpha_{{1x}}\alpha_{{1y}}\alpha_{{2x}}\alpha_{{2y}}}{4(\alpha_{{1x}}-\alpha_{{1y}})(\alpha_{{2x}}-\alpha_{{2y}})},
\eeq{4.19a}
which is in agreement with \eq{4.18c} when one sets $e=0$. Of course this equation generally will not have a solution for real $x_0$ if any of the polarizabilities are complex.

Additionally let us suppose that the polarizabilities $\alpha_{2x}$ and $\alpha_{2y}$ are very small, and $\alpha_{2x}\ne \alpha_{2y}$. Specifically let suppose
that $\alpha_{2y}=k\alpha_{2x}$, where $k$ is a fixed constant not equal to $1$. Then if the limit $\alpha_{2x}\to 0$ is taken {\it after}  the limit $e\to 0$, \eq{4.19}
implies  
\beq \BGb\to
\left[ \begin {array}{c} {\frac {\alpha_{{1x}}\alpha_{{1y}}}{\alpha_{{1
x}}-\alpha_{{1y}}}}\\ \noalign{\medskip}-{\frac {\alpha_{{1x}}\alpha_{{1y}
}x_{{0}}}{ \left( \alpha_{{1x}}-\alpha_{{1y}} \right) y_{{0}}}}
\\ \noalign{\medskip}0\\ \noalign{\medskip}0\end {array} \right].
\eeq{4.20}
By contrast if $\alpha_{2y}=k\alpha_{2x}$ , where $k$ is a fixed constant, and we take the limit $\alpha_{2x}\to 0$ directly in \eq{4.18}, keeping $e$ fixed and non-zero,
then we obtain
\beq \BGb\to  \left[ \begin {array}{c} -\alpha_{{1x}}\\ \noalign{\medskip}0
\\ \noalign{\medskip}0\\ \noalign{\medskip}0\end {array} \right].
\eeq{4.21}
Thus even a polarizable dipole at $(x_0,y_0)$ with very small polarizability can have a very large effect on the net dipole moments of the system if it is positioned
close to one of the characteristic lines where $e$ is small. Even though its polarizability is small the field exerted by this dipole on the dipole at the origin is still significant. We call this the {\it searchlight effect} since $\mu$ and thus the angle of the characteristic lines will depend on frequency, so by varying the frequency
and observing the net dipole moments of the system one may hope to detect something about the relative location of the polarizable dipoles, even though one of the polarizable dipoles, by itself, is difficult to detect.   The effect is illustrated in figures 2 and 3.

\begin{figure}[htbp]
\begin{center}
\scalebox{0.5}{\includegraphics{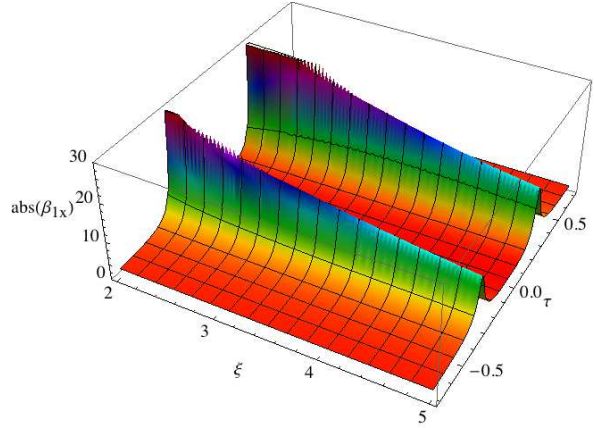}} 
\end{center}
\caption{Plot of the absolute value of the dipole amplitude $\Gb_{1x}$ near the line where $e\approx 0$ with
polarizabilities $\Ga_{1x}=2$, and $\Ga_{1y}=1$, $\Ga_{2x}=0.2$, and $\Ga_{2y}=0.1$ and parameter $c=0.01-i$.
We use rotated coordinates $\xi=(x_0+y_0)/\sqrt{2}$ and $\tau=(x_0-y_0)/\sqrt{2}$. Note the long-range
interaction.}
\label{fig:2}
\end{figure}

\begin{figure}[htbp]
\begin{center}
\scalebox{0.5}{\includegraphics{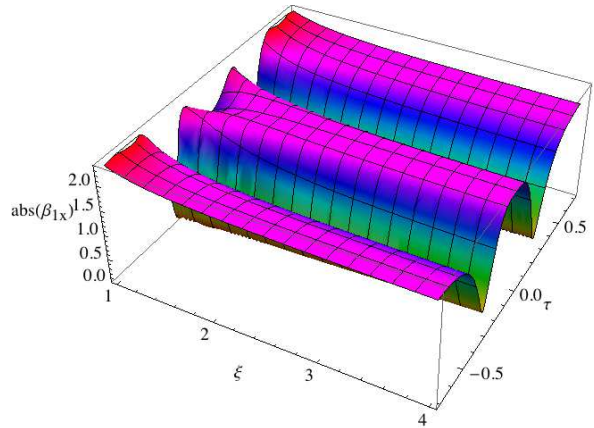}} 
\end{center}
\caption{Same as figure 2 but with polarizabilities $\Ga_{1x}=2$, and $\Ga_{1y}=1$, $\Ga_{2x}=0.1$, and $\Ga_{2y}=0.2$ and parameter $c=0.01-i$.}
\label{fig:3}
\end{figure}

The case $k=1$ when $\alpha_{2y}=\alpha_{2x}$ is rather special, but very interesting. In this case, with $e\to 0$ \eq{4.19} simplifies to
\beq \BGb \to
 \left[ \begin {array}{c}{2\,x_{{0}}}^{2}\\ \noalign{\medskip}-{
\frac {2\,{x_{{0}}}^{3}}{y_{{0}}}}\\ \noalign{\medskip}{\frac {2\,{x_{{0}
}}^{2} \left( \alpha_{{1x}}\alpha_{{1y}}-2\,{x_{{0}}}^{2}\alpha_{{1x}}+2
\,{x_{{0}}}^{2}\alpha_{{1y}} \right) }{\alpha_{{1x}}\alpha_{{1y}}}}
\\ \noalign{\medskip}-{\frac {2\,{x_{{0}}}^{3} \left( \alpha_{{1x}}
\alpha_{{1y}}-2\,{x_{{0}}}^{2}\alpha_{{1x}}+2\,{x_{{0}}}^{2}\alpha_{{1y}}
 \right) }{y_{{0}}\alpha_{{1y}}\alpha_{{1x}}}}\end {array} \right],
\eeq{4.22}
and this result does not depend on the magnitude of $\alpha_{2x}$. Remarkably, note that the magnitudes of the components of $\BGb$ increase as $x_0$ increases: recalling that
when $e=0$, $y_0=\pm x_0/(ic)$ we see that $\Gb_{1x}$ and $\Gb_{1y}$ increase as $x_0^2$, while $\Gb_{2x}$ and $\Gb_{2y}$ increase as $x_0^4$ when
$\Ga_{1x}\ne\Ga_{1y}$ and as $x_0^2$ when $\Ga_{1x}=\Ga_{1y}$. Thus the interaction increases the further the polarizable dipoles are apart!

\begin{figure}[htbp]
\begin{center}
\scalebox{0.5}{\includegraphics{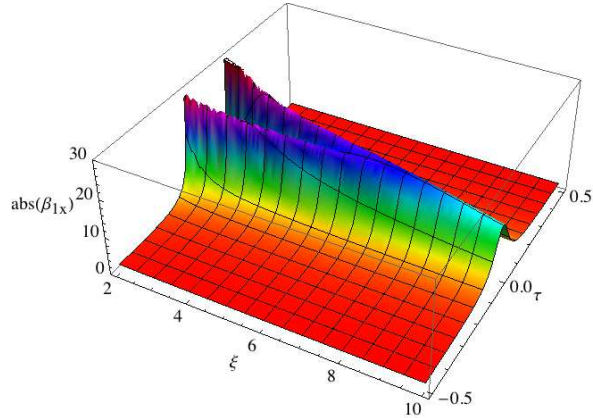}} 
\end{center}
\caption{Plot of the absolute value of the dipole amplitude $\Gb_{1x}$ near the line where $e\approx 0$ with
polarizabilities $\Ga_{1x}=2$, and $\Ga_{1y}=1$, $\Ga_{2x}=\Ga_{2y}=0.1$ and parameter $c=0.01-i$.
We use rotated coordinates $\xi=(x_0+y_0)/\sqrt{2}$ and $\tau=(x_0-y_0)/\sqrt{2}$. Note that the interaction
along the line $\tau=0$ first becomes stronger as $\xi$ increases, then weakens.}
\label{fig:4}
\end{figure}

To shed more light on this one can, using Maple, directly compute the right hand side of \eq{4.18} when $\alpha_{2y}=\alpha_{2x}$ to obtain
\beq \BGb=
 \left[ \begin {array}{c} {\frac { \left( -{e}^{2}+e\alpha_{{2x}}-2\,{x_{
{0}}}^{2}\alpha_{{2x}} \right) \alpha_{{1x}}}{{e}^{2}-\alpha_{{1x}}\alpha_{2
{x}}}}\\ \noalign{\medskip}-{\frac {2\,\alpha_{{1y}}x_{{0}} \left( e-{x
_{{0}}}^{2} \right) \alpha_{{2x}}}{ \left( {e}^{2}-\alpha_{{2x}}\alpha_{{1y
}} \right) y_{{0}}}}\\ \noalign{\medskip}{\frac {\alpha_{{2x}}f}
{{\left( {e}^{2}-\alpha_{{2x}}\alpha_{{1y}} \right)  \left( {e}^{2}-\alpha
_{{1x}}\alpha_{{2x}} \right)}}}
\\ \noalign{\medskip}-{\frac {2\,x_{{0}} \left( e-{x_{{0}}}^{2}
 \right) \alpha_{{2x}} g}{y_{{0}} \left( {e}^{2}-\alpha_{{2x}}\alpha_{{1y}} \right)  \left( {e}^{2}-\alpha
_{{1x}}\alpha_{{2x}} \right)}}\end {array} \right],
\eeq{4.23}
where
\beqa f &= & {e}^{3}\alpha_{{1x}}-2\,{x_{{0}}}^{2}\alpha_{{1x}}{e}^{2}-e\alpha_{{1x}}
\alpha_{{2x}}\alpha_{{1y}}+2\,{x_{{0}}}^{2}\alpha_{{1x}}\alpha_{{2x}}\alpha_
{{1y}}-{e}^{4} \nonum
& &+ 4\,\alpha_{{1x}}\alpha_{{2x}}e{x_{{0}}}^{2}-4\,\alpha_{{1x}}
\alpha_{{2x}}{x_{{0}}}^{4}+\alpha_{{2x}}\alpha_{{1y}}{e}^{2}-4\,{x_{{0}}}^{
2}\alpha_{{1y}}\alpha_{{2x}}e+4\,{x_{{0}}}^{4}\alpha_{{1y}}\alpha_{{2x}},
 \nonum
g & = &\alpha_{{1x}}{e}^{2}-\alpha_{{1y}}\alpha_{{1x
}}\alpha_{{2x}}-\alpha_{{1x}}\alpha_{{2x}}e+\alpha_{{1y}}\alpha_{{2x}}e+2\,
\alpha_{{1x}}\alpha_{{2x}}{x_{{0}}}^{2}-2\,{x_{{0}}}^{2}\alpha_{{1y}}\alpha
_{{2x}}.
\eeqa{4.24}

If the limit $e\to 0$ is taken in this expression we recover \eq{4.22}. On the other hand it is evident from \eq{4.23} that there are resonances when 
${e}^{2}$ equals $\alpha_{1x}\alpha_{2x}$ or $\alpha_{1y}\alpha_{2x}$, and that the resulting expression for $\BGb$ depends crucially on the ratio of the magnitude of $e^2$  to these two quantities. Also if $\Gn$ is non-zero the interaction decreases for sufficiently large separations of 
the polarizable dipoles. On the characteristic lines $x_0=\pm \Gm y_0$ we have $e\approx -2i\Gm\Gn y_0^2$. So for large $x_0=\pm \Gm y_0$ \eq{4.23} implies that, for example,
\beq \Gb_{1x}\approx \frac{x_0^2\Ga_{2x}\Ga_{1x}}{2\Gm^2\Gn^2 y_0^4}=\frac{\Ga_{2x}\Ga_{1x}}{2\Gn^2 y_0^2},
\eeq{4.25}
which goes to zero as $1/y_0^2$ as $y_0\to \infty$. Figure 4 shows how,  as the separation increases, the interaction along the characteristic line first increases, then decreases.

\section*{Acknowledgements}
Graeme Milton is thankful to Ben Eggleton, CUDOS and the University of Sydney for the provision of 
office space during his visit there and to the
National Science Foundation for support through grant DMS-1211359. R.C. McPhedran acknowledges the support of the Australian Research Council through its Discovery Grant Scheme. The Centre for Ultrahighbandwidth Devices for Optical Systems (CUDOS) is an
ARC Centre of Excellence (Project No. CE110001018).

\ifx \bblindex \undefined \def \bblindex #1{} \fi\ifx \bblindex \undefined \def
  \bblindex #1{} \fi


\end{document}